# Controls of Atmospheric Methane on Early Earth and Inhabited Earth-like Terrestrial Exoplanets

Aika Akahori[1], Yasuto Watanabe[1,2,*], Eiichi Tajika[1]

## Abstract

Methane ($CH_4$) is a primarily biogenic greenhouse gas. As such, it represents an essential biosignature to search for life on exoplanets. Atmospheric $CH_4$ abundance on Earth-like inhabited exoplanets is likely controlled by marine biogenic production and atmospheric photochemical consumption. Such interactions have been previously examined for the case of the early Earth where primitive marine ecosystems supplied $CH_4$ to the atmosphere, showing that the atmospheric $CH_4$ response to biogenic $CH_4$ flux variations is nonlinear, a critical property when assessing $CH_4$ reliability as a biosignature. However, the contributions of atmospheric photochemistry, metabolic reactions, or solar irradiance to this nonlinear response are not well understood. Using an atmospheric photochemical model and a marine microbial ecosystem model, we show that production of hydroxyl radicals from water vapor photodissociation is a critical factor controlling the atmospheric $CH_4$ abundance. Consequently, atmospheric $CH_4$ partial pressure ($pCH_4$) on inhabited Earth-like exoplanets orbiting Sun-like stars (F-, G-, and K-type stars) would be controlled primarily by stellar irradiance. Specifically, irradiance at wavelengths of approximately 200–210 nm is a major controlling factor for atmospheric $pCH_4$ when the carbon dioxide partial pressure is sufficiently high to absorb most stellar irradiance at 170–200 nm. Finally, we also demonstrated that inhabited exoplanets orbiting near the outer edge of K-type stars' habitable zones are better suited for atmospheric $pCH_4$ buildup. Such properties will valuably support future detection of life signatures.

[1] Department of Earth and Planetary Science, The University of Tokyo, Hongo 7-3-1, Bunkyo-ku, Tokyo, Japan. 113-0033
[2] Meteorological Research Institute, Japan Meteorological Agency, Nagamine 1-1, Tsukuba, Ibaraki, Japan. 305-0052
[*] Corresponding Author yasuto.watanabe.wess@gmail.com

1. INTRODUCTION

Solar luminosity on early Earth during the Hadean (> 4.0 Ga) and Archean (4.0–2.5 Ga) was ~70–80% of its current value (Gough 1981). Even under such young Sun conditions, early Earth was maintained habitable by the presence of liquid water at its surface (Sagan and Mullen 1972; Mojzsis et al. 2001; Wilde et al. 2001; Feulner 2012; Catling and Zahnle 2020), ensured by elevated concentrations of greenhouse gases, the major contributors to early-Earth warming mechanisms (Sagan and Mullen 1972; Walker et al. 1981; Sagan and Chyba 1997; Haqq-Misra et al. 2008; Goldblatt and Zahnle 2011; Wordsworth and Pierrehumbert 2013). Carbon dioxide ($CO_2$) is one of the most fundamental greenhouse gasses in planetary atmospheres, with estimated early-Earth concentrations approximately 100 to 1000 times larger than the preindustrial reference value (Tajika and Matsui 1992; Kasting 1993; Krissansen-Totton et al. 2018; Isson and Planavsky 2018; Kadoya et al. 2020). Methane ($CH_4$), the second most abundant greenhouse gas in the current atmosphere, was initially produced and emitted into the early-Earth atmosphere by a primitive marine microbial ecosystem (Kiehl and Dickinson 1987; Pavlov et al. 2001; Kasting 2005; Haqq-Misra et al. 2008; Ozaki et al. 2018). The first living organisms emerged during the Hadean or early Archean (4.0–3.5 Ga), less than 1 Gyr after the Earth formed (Lazcano and Miller 1994; Chyba and McDonald 1995; Ueno et al. 2006; Dodd et al. 2017; Tashiro et al. 2017; Pearce et al. 2018). Moreover, methanogens, microorganisms that produce $CH_4$ by decomposing organic matter, probably emerged nearly simultaneously, during the early to middle Archean (Battistuzzi et al. 2004; Ueno et al. 2006; Wolfe and Fournier 2018). Although $CH_4$ is less abundant than $CO_2$ on current Earth, it is also a considerably more potent greenhouse gas. Therefore, the emergence of life may have fundamentally influenced the early atmospheric composition. Theoretical studies have suggested that biogenic $CH_4$ accumulated in the early atmosphere, warming the Earth surface (Pavlov et al. 2001; Ozaki et al. 2018).

Relationships between the atmosphere and marine microbial ecosystem activity on early Earth are particularly interesting in relation to the investigation of exoplanetary atmospheres. Because early $CH_4$ sources, as on Earth currently, were probably exclusively biological, $CH_4$ is also considered a potentially strong biosignature detectable from inhabited Earth-like exoplanets (Schindler and Kasting 2000; Kasting and Catling 2003; Catling et al. 2018; Wogan and Catling 2020; Wogan et al. 2020; Sauterey et al. 2020; Krissansen-Totton et al. 2022; Thompson et al. 2022). Concentrations of atmospheric biogenic gases respond nonlinearly to reducing-gas outgassing flux through changes in atmospheric photochemical reaction rates (Pavlov et al. 2003; Kasting 2005; Ozaki et al. 2018; Ranjan et al. 2022; Watanabe et al. 2023b). Therefore, understanding the response of atmospheric photochemical reactions to biological activity is essential to understand early-Earth evolution of the atmosphere and biosphere and to assess the potential role of $CH_4$ as a biosignature on exoplanets.

Using a coupled model that couples atmospheric photochemistry with a primitive marine microbial ecosystem (Figure 1), a previous study demonstrated that the high atmospheric $CH_4$ partial pressure ($pCH_4$) required to sustain warm climate conditions, within an estimated atmospheric $CO_2$ level ($pCO_2$) range representative of the mid-Archean, could be produced by the nonlinear response of atmospheric $pCH_4$ to primitive marine biospheric activities (Figure 1) (Ozaki et al. 2018). The primary production in the primitive marine microbial ecosystem was mainly due to anoxygenic photoautotrophs (AP), organisms that perform photosynthesis without producing oxygen using electron donors such as hydrogen ($H_2$) and ferrous iron (Fe(II)) (Kharecha et al. 2005; Canfield et al. 2006; Ozaki et al. 2018; Ward et al. 2019). This leads to an increase in biogenic $CH_4$ production rate and atmospheric $p$CH$_4$ by enhancing the activity of methanogenesis. Another plausible source of CH$_4$ on early Earth was the activity of chemoautotrophs, which utilize carbon monoxide (CO) as an electron donor (CO-consuming chemoautotroph; CO-C) (Ragsdale 2004; Kharecha et al. 2005; Ozaki et al. 2018; Schwieterman et al. 2019; Watanabe et al. 2023b). However, possible causes for nonlinearity of the atmospheric $p$CH$_4$ response, in terms of interactions between metabolic pathways and atmospheric composition, have not been systematically assessed. Specifically, the roles of the nonlinear patterns of CH$_4$ and the influence of solar irradiance on atmospheric photochemical reactions, also critical to analyzing exoplanetary atmospheres, remain unclear. In this study, we use an atmospheric photochemical model and a marine microbial ecosystem model to investigate possible factors controlling CH$_4$ abundance in the early-Earth atmosphere and on Earth-like inhabited exoplanets in orbit around F-, G-, and K-type central stars.

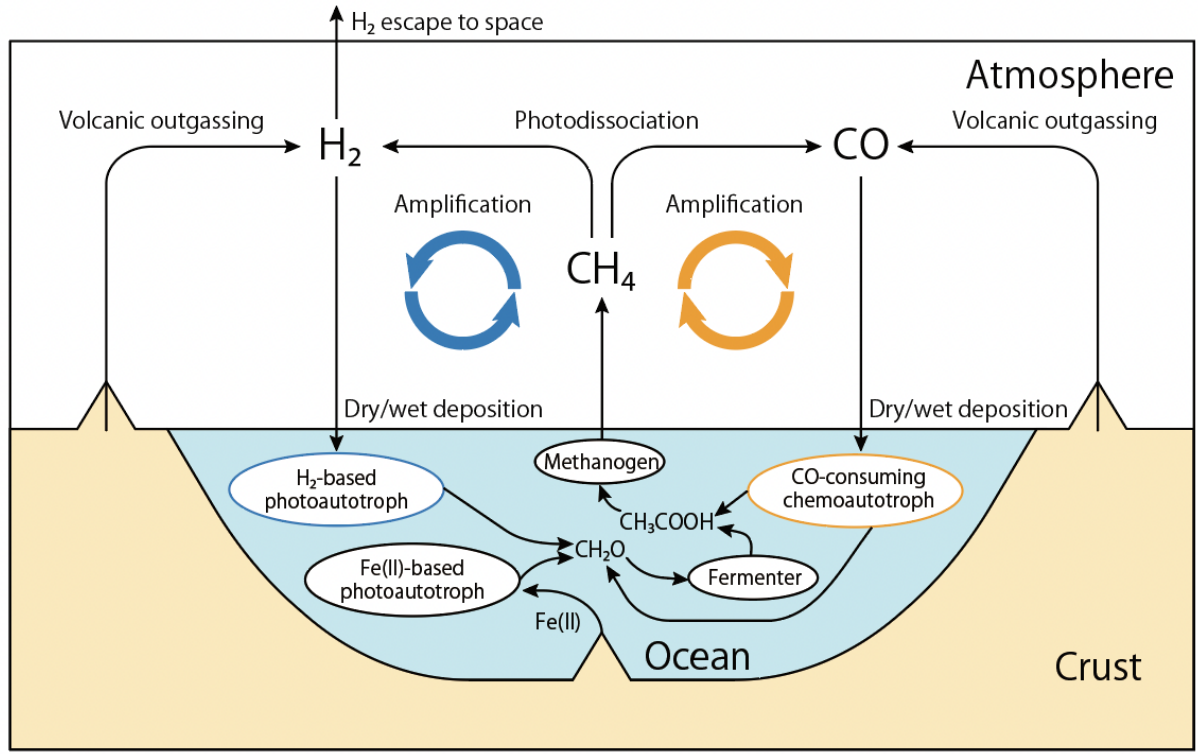

Figure 1. Schematic illustration of the interactions of atmospheric photochemistry and marine microbial ecosystem on early Earth based on Ozaki et al. (2018).

2. METHODS

2.1. Coupled Model for Atmospheric Photochemistry and Marine Microbial Ecosystems

We used the coupled model for atmospheric photochemistry and marine microbial ecosystems that includes the 1D photochemical model "Atmos" (Arney et al. 2016, 2017, 2018) and the primitive marine microbial ecosystem model (Ozaki et al. 2018; Watanabe et al. 2023b). The photochemical model version; lower boundary conditions; absorption cross-sections (Figure A1) (Ranjan et al. 2020); and vertical profiles of atmospheric number density, temperature, and eddy diffusion coefficient were identical to those used in Watanabe et al. (2023b).

The ecosystem model is similar to the earlier model of Ozaki et al. (2018), but with an improved formulation (Watanabe et al. 2023b). Biological reactions implemented in the model are summarized in Table 1. They describe the activity of $H_2$-based and Fe(II)-based AP (hereafter $H_2$-AP and Fe-AP, respectively), CO-C, fermenters, and methanogens. In this study, we assumed two types of primitive marine microbial ecosystems (Ozaki et al., 2018), differing by the omission (Type I ecosystem) or inclusion (Type II ecosystem) of Fe-AP activity and ferric iron (Fe(III)) reduction (Table 1). The $CH_4$ flux from the ocean into the atmosphere, $\Phi_\uparrow(CH_4)$, is expressed as (Kharecha et al. 2005; Ozaki et al. 2018; Watanabe et al. 2023b):

$$\Phi_\uparrow(CH_4) = \frac{1}{2}(1 - \beta_{bur}) \cdot F_{npp} + \frac{1-g_{co}}{4-2g_{co}} \Phi_\downarrow(CO) + \Phi_\downarrow(CH_4) \qquad \text{(Type I)},$$

$$= \frac{1}{2}(1 - \beta_{bur}) \cdot F_{npp} + \frac{1-g_{co}}{4-2g_{co}} \Phi_\downarrow(CO) - \frac{\gamma_{red}}{8} \Phi_\uparrow(Fe(II)) + \Phi_\downarrow(CH_4) \quad \text{(Type II)}, \quad (1)$$

where $\Phi_\downarrow(CO)$ and $\Phi_\downarrow(CH_4)$ are the total deposition rates (sum of the dry and wet deposition rates) of CO and $CH_4$, respectively; $\beta_{bur}$ is the organic carbon burial efficiency, $F_{npp}$ is the net primary productivity of the anaerobic marine ecosystem; $g_{co}$ is the growth yield of CO-C; $\gamma_{red}$ is the iron reduction efficiency, and $\Phi_\uparrow(Fe(II))$ is the rate of Fe(II) upwelling from the deep ocean to the euphotic zone. The standard parameters are summarized in Tables 2 and 3. We assumed that the primitive marine ecosystem productivity is limited by the electron donors supply rate. Thus, $F_{npp}$ is expressed as:

$$F_{npp} = \frac{1}{2}\Phi_\downarrow(H_2) + \frac{g_{co}}{4-2g_{co}}\Phi_\downarrow(CO) \qquad \text{(type I)}$$

$$= \frac{1}{2}\Phi_\downarrow(H_2) + \frac{1}{4}\Phi_\uparrow(Fe^{II}) + \frac{g_{co}}{4-2g_{co}}\Phi_\downarrow(CO) \qquad \text{(type II)} \qquad (2)$$

where $\Phi_\downarrow(H_2)$ is the $H_2$ surface removal flux (sum of the dry and wet deposition fluxes).

**Table 1.** List of biological reactions. $g_{co}$ is the growth yield of CO-consuming chemoautotroph. Note, the reaction B6 is included in the model but its activity is set to zero in this study.

| # | Reaction | Reaction | Type I | Type II |
|---|---|---|---|---|
| B1 | $H_2$-based anoxygenic photosynthesis | $2H_2 + CO_2 + hv \rightarrow CH_2O + H_2O$ | ✓ | ✓ |
| B2 | Fe(II)-based anoxygenic photosynthesis | $4Fe^{2+} + 11H_2O + CO_2 + hv \rightarrow CH_2O + 8H^+$ | | ✓ |
| B3 | CO-consuming chemoautotroph | $(4 - 2g_{co}) \cdot CO + (2 - g_{co}) \cdot H_2O$ $\rightarrow g_{co} \cdot CH_2O + (2 - g_{co}) \cdot CO_2 + (1 - g_{co}) \cdot CH_3COOH$ | ✓ | ✓ |
| B4 | Fermenter | $2CH_2O \rightarrow CH_3COOH$ | ✓ | ✓ |
| B5 | Methanogen | $CH_3COOH \rightarrow CH_4 + CO_2$ | ✓ | ✓ |
| B6 | Fe(III) reduction | $CH_3COOH + 8Fe(OH)_3 + 16H^+ \rightarrow 8Fe^{2+} + 22H_2O + 2CO_2$ | | ✓ |

Once the model reaches steady state, the net oxidizing and reducing power supply rates into the ocean–atmosphere system become balanced. The model global redox budget then becomes (Holland 1984; Canfield et al. 2006; Ozaki et al. 2018; Watanabe et al. 2023b):

$$F_{volc,red} + 0.5 F_{Fe(III)} = F_{esc,H2} + 2F_{bur,oc} + \Sigma\Delta\Phi_{red} \qquad (3)$$

where $F_{volc,red}$ is the volcanic outgassing rate for reduced gases, $F_{Fe(II)}$ is the Fe(III) deposition rate from the ocean, $F_{esc,H2}$ is the H$_2$ escape rate to space, $F_{bur,oc}$ is the organic carbon burial rate from the ocean, and $\Sigma\Delta\Phi_{red}$ is a residual term that represents the net removal rate of reducing power (converted into units of H$_2$ reducing power) via deposition of atmospheric chemical species (Watanabe et al. 2023b). The left-hand side in equation (3) represents the reducing power supply rate (oxidizing power removal rate). Conversely, the right-hand side represents the oxidizing power supply rate (reducing power removal rate). The term $F_{volc,red}$ is expressed as the sum of H$_2$, CO, and hydrogen sulfide (H$_2$S) outgassing fluxes ($\Phi_\uparrow$(H$_2$), $\Phi_\uparrow$(CO), and $\Phi_\uparrow$(H$_2$S), respectively):

$$F_{volc,red} = \Phi_\uparrow(H_2) + \Phi_\uparrow(CO) + 3\Phi_\uparrow(H_2S), \qquad (4)$$

where the multiplying factor for $\Phi_\uparrow$(H$_2$S) is required for unit conversion to the reducing power equivalent to H$_2$. The outgassing rates in all simulations are summarized in Tables 2 and 3. We did not consider CH$_4$ abiotic outgassing flux because it is much smaller than the biogenic flux (Thompson et al. 2022). In these conditions, the atmospheric redox budget becomes:

$$F_{volc,red} + 4\Phi_\uparrow(CH_4) = F_{esc,H2} + \Sigma\Delta\Phi_{red} \qquad (5)$$

From equation (5), we infer that the increase in reducing power supply rate to the atmosphere provided by biogenic CH$_4$ is compensated (removed from the atmosphere) through both escape of H$_2$ to space and deposition of reducing chemical species from the atmosphere into the ocean.

**Table 2.** List of the standard parameters adopted in this study.

| Parameter | Label | Value | Unit | Remarks and Reference |
|---|---|---|---|---|
| Atmospheric $pCO_2$ | $pCO_2$ | 0.03 (Archean Earth)<br>0.1 (F-, G-, and K-type) | bar | – |
| Stellar irradiance (present Earth = 1) | $S^*$ | 0.8 (Archean Earth)<br>1.0 (F-, G-, and K-type) | – | (Gough, 1981) |
| Volcanic outgassing rate of $H_2$ | $\Phi_{out}(H_2)$ | 2.7 | Tmol $H_2$ yr$^{-1}$ | (Catling and Kasting 2017; Ozaki et al. 2018) |
| Volcanic outgassing rate of $H_2S$ | $\Phi_{out}(H_2S)$ | 0.0945 | Tmol S yr$^{-1}$ | (Wogan and Catling 2020; Watanabe et al. 2023b) |
| Fraction of Fe(III) decomposed by Fe(III)-reducing bacteria | $r_{fered}$ | 0 | – | (Ozaki et al., 2018) |
| Burial efficiency of organic carbon | $\beta_{bur}$ | 0.02 | – | (Betts and Holland 1991; Ozaki et al. 2018) |
| Growth yield of the CO-consuming chemoautotroph | $g_{co}$ | 0.1 | – | (Kharecha et al. 2005; Watanabe et al. 2023b) |
| Unit conversion factor from cm$^{-2}$ s$^{-1}$ to Tmol yr$^{-1}$ | – | 270 | – | – |

**Table 3.** List of numerical experiments conducted in this study.

| Experiment name | $\Phi_\uparrow(H_2)$ (cm$^{-2}$ s$^{-1}$) | $\Phi_\uparrow(CO)$ (cm$^{-2}$ s$^{-1}$) | $\Phi_\uparrow(Fe(II))$ (cm$^{-2}$ s$^{-1}$) | $\Phi_\uparrow(CH_4)$ (cm$^{-2}$ s$^{-1}$) | $pCO_2$ (bar) | Stellar Spectrum | Remarks |
|---|---|---|---|---|---|---|---|
| *ExpH2outgassTypeI* | $1\times10^8$–$3\times10^{11}$ | 0 | N/A | N/A | 0.03 | Early Sun (2.8 Ga) G-type (Sun) | No Fe(II)-based anoxygenic photoautotroph Ozaki et al. (2018) equivalent |
| *ExpH2outgassTypeII* | $1\times10^8$–$3\times10^{11}$ | 0 | $3\times10^{11}$ | N/A | 0.03 | Early Sun (2.8 Ga) | Ozaki et al. (2018) equivalent |
| *ExpCOoutgassTypeI* | $1\times10^{10}$ | $1\times10^8$–$3\times10^{11}$ | N/A | N/A | 0.03 | Early Sun (2.8 Ga) | No Fe(II)-based anoxygenic photoautotroph |
| *ExpCH4flux* | $1\times10^{10}$ | 0 | N/A | $1\times10^6$–$9\times10^{11}$ | 0.001 0.003 0.01 0.03 0.1 0.342 1.2 | Early Sun (2.8 Ga) | – |
| *ExpCH4fluxG2V* | $1\times10^{10}$ | 0 | N/A | $1\times10^6$–$9\times10^{11}$ | 0.001 0.01 0.1 | G-type (Sun) | – |
| *ExpCH4fluxF2V* | $1\times10^{10}$ | 0 | N/A | $1\times10^6$–$9\times10^{11}$ | 0.001 0.01 0.1 | F-type (σ-Bootis) | – |
| *ExpCH4fluxK2V* | $1\times10^{10}$ | 0 | N/A | $1\times10^6$–$9\times10^{11}$ | 0.001 0.01 0.1 | K-type (ε-Eridani) | – |
| *ExpCH4flux200nm* | $1\times10^{10}$ | 0 | N/A | $1\times10^6$–$9\times10^{11}$ | 0.1 | G-type (Sun) but irradiance at 200–210 nm is scaled by 0.1, 0.5, 1, 2, 10, or 40. | – |
| *ExpIrradiance* | $1\times10^{10}$ | 0 | N/A | $1\times10^9$ $3\times10^9$ $1\times10^{10}$ $3\times10^{10}$ $1\times10^{11}$ | 0.1 | G-type (Sun) scaled by 0.2–1.2 | – |

2.2. Experimental Setup

We first conducted two sets of simulations similar to those of Ozaki et al. (2018): one set for a Type I ecosystem including $H_2$-AP, CO-C, and decomposers, but without Fe-AP (Table 3, *ExpH2outgassTypeI*) and another set for a Type II ecosystem including $H_2$-AP, Fe-AP, CO-C, and decomposers (Table 3, *ExpH2outgassTypeII*). In each set, we conducted simulations with respect to a different $H_2$ outgassing rate value (from $1\times10^8$ to $3\times10^{11}$ cm$^{-2}$ s$^{-1}$). In addition, we also conducted an additional sensitivity test to the CO outgassing flux for the Type I ecosystem (Table 3, *ExpCOoutgassTypeI*). These simulations were conducted with a fixed atmospheric $pCO_2$ of 0.03 bar (Table 3), a reasonable value for late Archean conditions (Kanzaki and Murakami 2015; Ozaki et al. 2018).

Then, we conducted additional sensitivity tests to assess the influence of the $CH_4$ flux from the ocean into the atmosphere (from $1\times10^6$ to $9\times10^{11}$ cm$^{-2}$ s$^{-1}$) on the atmospheric composition (Table 3, *ExpCH4flux*). These simulations used the same photochemical model decoupled from the ecosystem model, allowing us to assess the atmospheric response to a broader range of $CH_4$ fluxes than with using the coupled model. This set of simulations was repeated for different values of atmospheric $pCO_2$ (0.001, 0.003, 0.01, 0.03, 0.1, 0.342, and 1.2 bar). Total atmospheric pressure was adjusted when atmospheric $pCO_2$ was larger than 0.2 bar (Watanabe and Ozaki 2023). The $H_2$ outgassing rate was set to $10^{10}$ cm$^{-2}$ s$^{-1}$. Solar luminosity was set to ~0.8 times its current value, representative of "young Sun" conditions (2.8 Ga).

Finally, we conducted three sets of simulations for central stars of F, G, and K types (Table 3, *ExpCH4fluxF2V*, *ExpCH4fluxG2V*, and *ExpCH4fluxK2V*, respectively), with three settings in each case for the atmospheric $pCO_2$ values of 0.001, 0.01, and 0.1 bar. The adopted irradiance spectra (Figure 2) were those of $\sigma$ Bootis for F-type stars (Segura et al. 2003), the current Sun for G-type stars (Thuillier et al., 2004), and $\varepsilon$ Eridani for K-type stars (Segura et al. 2003). All spectra were scaled so that stellar irradiance at the top of the exoplanetary atmosphere is equal to the solar value on current Earth. We conducted further simulations for a G-type star (current Sun), but with (1) irradiance at wavelengths within ~200–210 nm scaled successively by a factor of 0.1, 0.5, 1, 2, 10, and 40 (Table 3, *ExpCH4flux200nm*); and (2) with finer solar irradiance scaling at all wavelengths (factor of 0.2–1.2; Table 3, *ExpIrradiance*). The irradiance scaling range in (2) is sufficient to represent expected conditions within the habitable zone of G-type stars, defined as the circumstellar region where a terrestrial planet with an atmospheric composition including mainly $CO_2$, water vapor ($H_2O$), and nitrogen ($N_2$) can sustain liquid water at its surface (Kasting et al. 1993; Selsis et al. 2007; Kopparapu et al. 2013, 2014). For the current Sun, the habitable zone limits are estimated to be ~0.99–1.70 au, corresponding to stellar irradiances of ~1.01–0.35 times that of Earth (Kopparapu et al. 2013). Both sets of simulations were repeated for five distinct $CH_4$ flux values ($1\times10^9$, $3\times10^9$, $1\times10^{10}$, $3\times10^{10}$, and $1\times10^{11}$ cm$^{-2}$ s$^{-1}$).

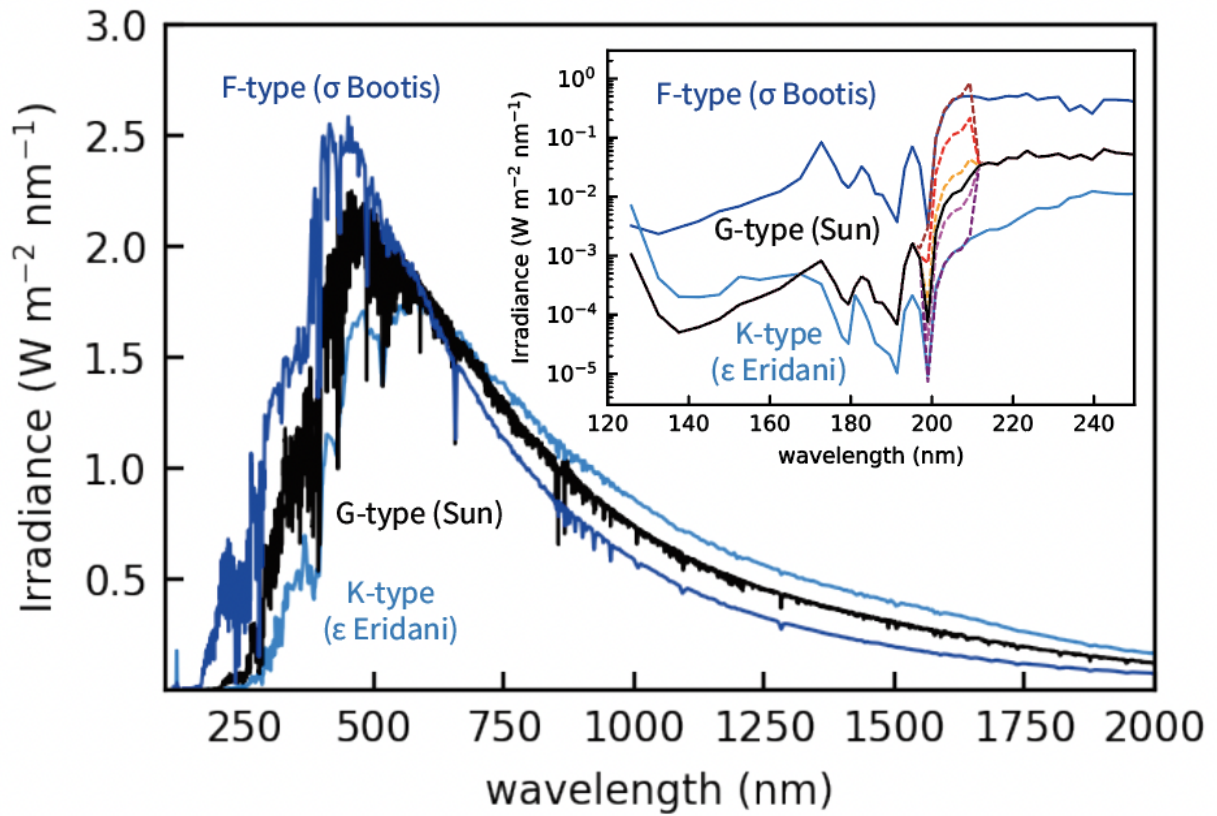

Figure 2. Spectrum of the radiation of the Sun-like stars (F-, G-, and K-type stars) employed in this study (blue, black, and light blue lines, respectively) (Segura et al., 2003; Thuillier et al., 2004). This original spectrum is translated into the model grid in wavelength and used as an input of the model. The upper-right window shows the ultraviolet region of the radiation spectrum that is translated into the model wavelength grid. The solid lines are the input spectrum of F-, G-, and K-type stars (blue, black, and light blue lines, respectively). The dashed lines represent the spectra of the G-type star, which are scaled 0.1, 0.5, 2, 10, and 40 times at 200–210 nm (purple, light purple, orange, red, and brown lines, respectively).

3. RESULTS

3.1. Global Redox Budget and CH$_4$ Amplification on Early Earth

Simulation results reproducing the study of Ozaki et al. (2018) are shown in Figure 3. The current coupled model version correctly reproduced the nonlinear response of atmospheric $p$CH$_4$ to H$_2$ outgassing rate variations. As discussed by Ozaki et al. (2018), calculated atmospheric $p$CH$_4$ was larger than 0.001 bar for Type II ecosystems (active Fe-AP; red lines in Figure 3), independently of the H$_2$ outgassing rate, but depended strongly on it for Type I ecosystems (black lines in Figure 3). This indicates nonlinear amplification of atmospheric $p$CH$_4$ for high Fe-AP activity, easily explained by analyzing the global redox budget. In Figures 3b, 3d, and 3f, the result is shown as a function of the influx of reducing power (equation (2), left-hand side). For Type II ecosystems, the electron dolors supply rate was always larger than 40 Tmol H$_2$ eq. yr$^{-1}$ because of the high Fe(II) upwelling rate (~80 Tmol Fe yr$^{-1}$; red line in Figure 3b). This clearly demonstrates that the influx of reducing power (i.e., global redox budget) controls the primitive marine biosphere activity, thus atmospheric $p$CH$_4$. When the influx of reducing power exceeds ~60 Tmol H$_2$ eq. yr$^{-1}$, CH$_4$ decreases slightly, whereas atmospheric $p$CH$_4$ increases. For a reducing power influx higher than this threshold, hydrocarbon haze, composed of hydrocarbon aerosols, forms in the atmosphere and affect CH$_4$ chemistry (Sagan and Chyba 1997; Pavlov et al. 2001; Arney et al. 2016). However, investigating hydrocarbon haze requires the consideration of temperature profile changes (Arney et al. 2016, 2017), so it is beyond the scope of this study. For this reason, we analyze the response of atmospheric $p$CH$_4$ when the atmospheric haze layer is not formed.

For simulated Type I ecosystems, increased H$_2$ outgassing flux (i.e., influx of reducing power) enhanced the H$_2$-AP activity. Consequently, both the biogenic CH$_4$ production rate and the atmospheric $p$CH$_4$ increases (Figure 3b). This implies that early-Earth atmospheric $p$CH$_4$ was determined by interactions between the atmosphere and the marine biosphere in response to an influx of reducing power to the ocean–atmosphere system. For this reason, in Type-I ecosystems, we focus on the relationships between the influx of reducing powers, $\Phi_\uparrow(CH_4)$, and atmospheric $pCH_4$.

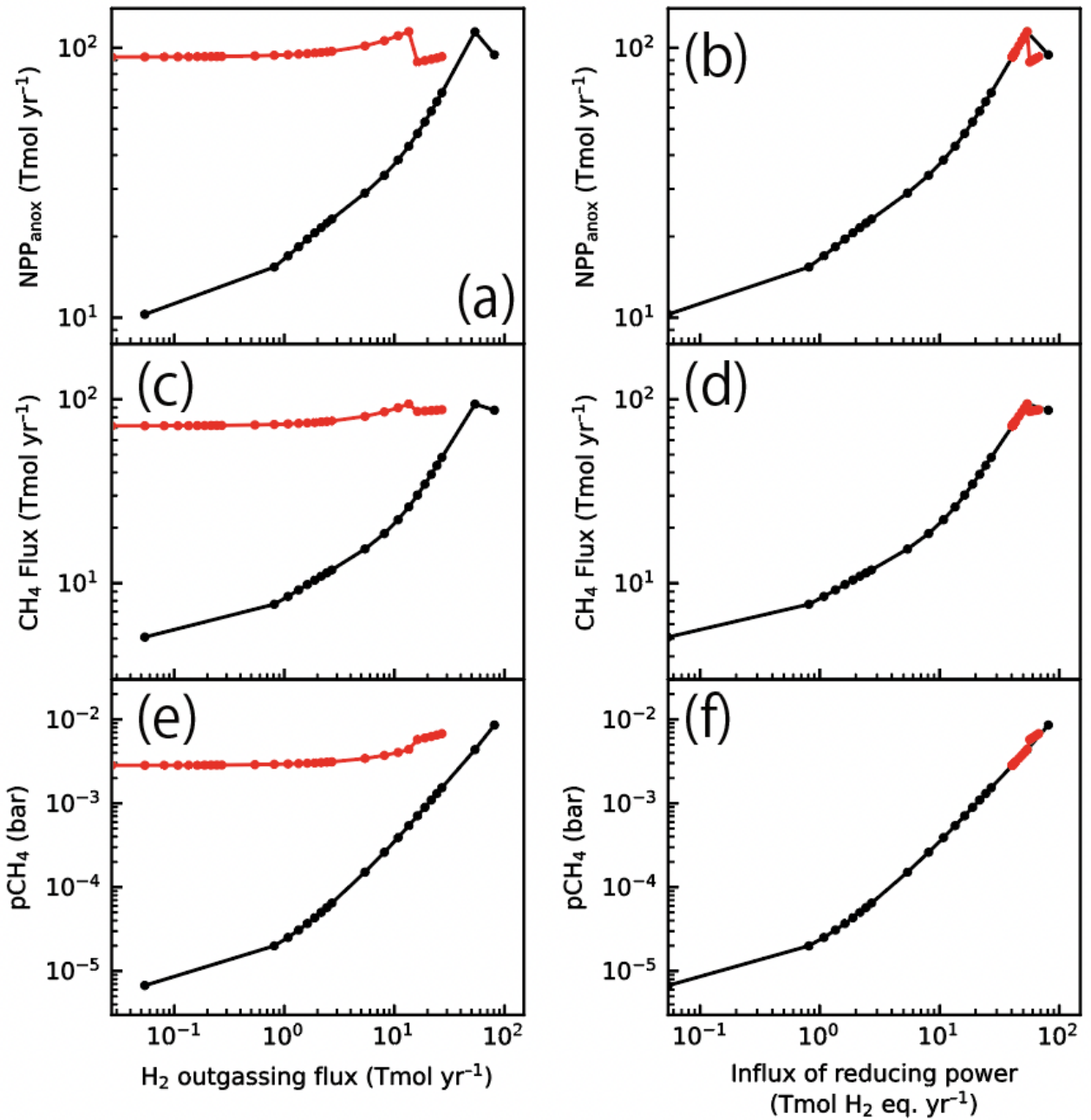

Figure 3. Response of primary productivity (a, b), biogenic $CH_4$ flux (c, d), and atmospheric $pCH_4$ in response to the changes in the outgassing flux of $H_2$ (a, c, e) and the influx of reducing power (left-hand side of equation (5)) (b, d, f), which are equivalent to the results in Ozaki et al. (2018). The black lines represent the result with $H_2$-based anoxygenic photoautotrophs (APs) but without Fe(II)-based APs (*ExpH2outgassTypeI* in Table 3). The red lines represent the result with $H_2$-using AP and Fe(II)-based AP, assuming the Fe(II) upwelling rate is approximately 80 Tmol Fe yr$^{-1}$ (*ExpH2outgassTypeII* in Table 3).

3.2. Response of the Biogenic $CH_4$ Supply Rate to $H_2$ and CO Outgassing Rate Variations

$CH_4$ flux amplification related to reducing gas outgassing (i.e., the influx of reducing power) occurs through two main pathways: recycling of atmospheric $H_2$ and CO by $H_2$-AP and CO-C, respectively (Figure 1). Responses of atmospheric $pCH_4$, CO partial pressure ($pCO$), and $H_2$ partial pressure ($pH_2$) to $\Phi_\uparrow(CH_4)$ are illustrated in Figure 4. Atmospheric $pCH_4$ and $pCO$ increase nonlinearly with increasing $\Phi_\uparrow(CH_4)$, whereas $pH_2$ exhibits a nearly linear dependence on the $CH_4$ flux (Figure 4c). When the $CH_4$ flux increases, the CO consumption flux via reaction with hydroxyl (OH) radicals (reaction (R1)) decreases rapidly below approximately 40 km (Figure 5b and 5c).

$$CO + OH \rightarrow CO_2 + H. \qquad (R1)$$

This decrease is explained by OH consumption by biogenic $CH_4$ when the $H_2$ outgassing rate is high (Figure 5e):

$$CH_4 + OH \rightarrow CH_3 + H_2O. \qquad (R2)$$

This causes a sharp increase in the wet and dry deposition rate of CO to the ocean and in the activity of CO-C. As a result, the $CH_4$ production rate increases, further enhancing CO production. We repeated this simulation with different CO outgassing fluxes to confirm the dependence of the $H_2$ and CO budgets on volcanic gas composition (Table 3, *ExpCOoutgassTypeI*). Results were similar to those obtained for different $H_2$ outgassing fluxes (Figures 4b and 4c). Therefore, the nonlinear responses of both the electron donors supply rate and $\Phi_\uparrow(CH_4)$ were primarily caused by abundance variations of CO, $CH_4$, and OH radicals.

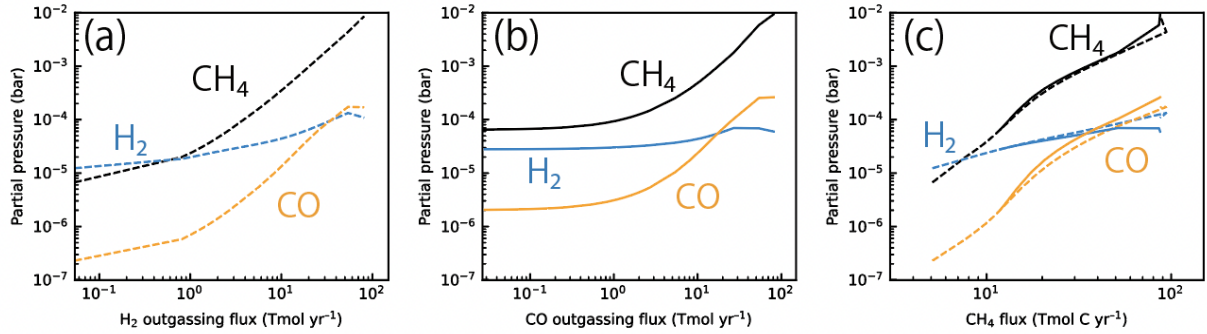

Figure 4. Response of the atmospheric $pCH_4$, $pH_2$, and $pCO$ (black, blue, and orange lines, respectively) to the change in $H_2$ outgassing flux (a), CO outgassing flux (b), and the biogenic $CH_4$ flux (c). Calculations are conducted with atmospheric $pCO_2$ of 0.03 bar (*ExpH2outgassTypeI* and *ExpCOoutgassTypeI* in Table 3). The dashed line represents the results with different $H_2$ outgassing fluxes. The solid line represents the results with different CO outgassing fluxes.

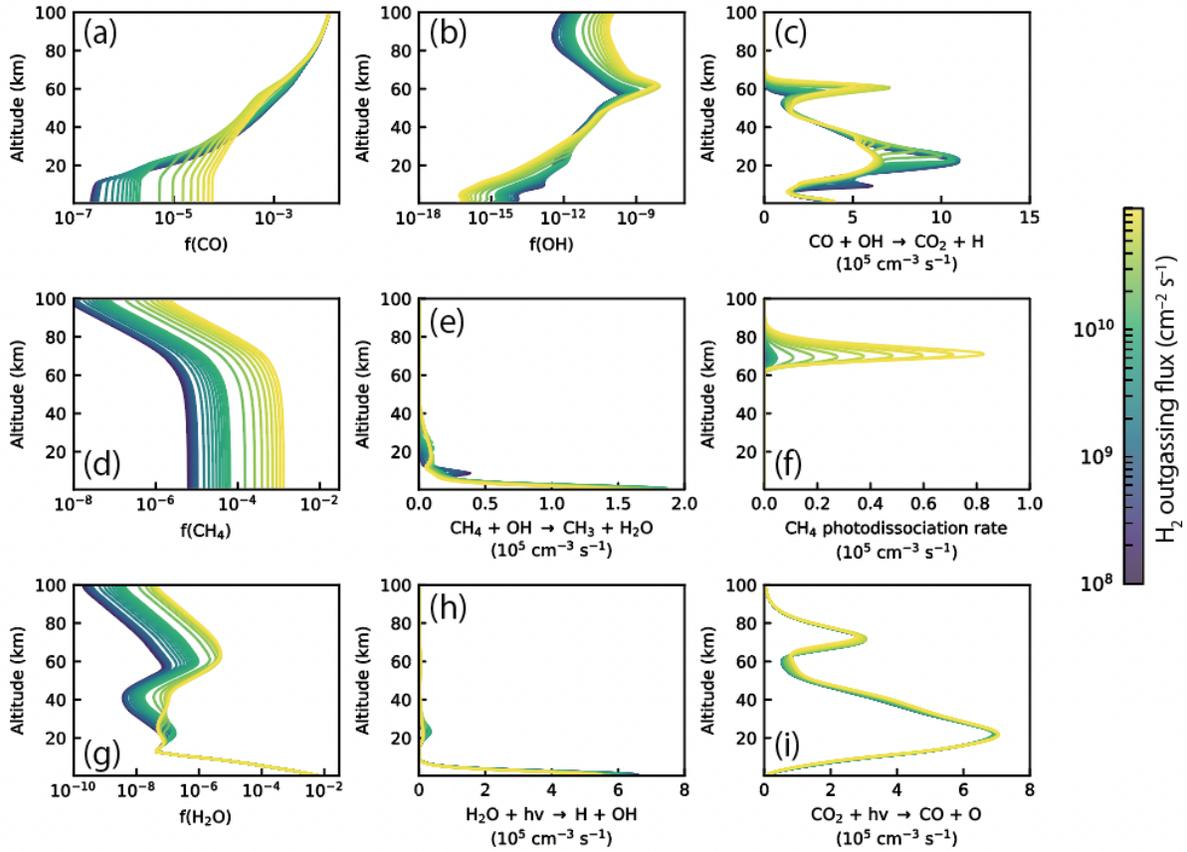

Figure 5. Vertical profiles of the atmospheric components and the reaction rates. Vertical profiles of (a) CO mixing ratio, (b) OH radical mixing ratio, (c) reaction rate of CO + OH → $CO_2$ + H, (d) $CH_4$ mixing ratio, reaction rate of (e) $CH_4$ + OH → $CH_3$ + $H_2O$, (f) photodissociation rate of $CH_4$, (g) $H_2O$ mixing ratio, (h) photodissociation rate of $H_2O$, and (i) photodissociation rate of $CO_2$. Calculations are conducted with $pCO_2$ of 0.03 bar (*ExpH2outgassTypeI* in Table 3). The line colors represent different $H_2$ outgassing rates.

3.3. Response of Atmospheric $p$CH$_4$ to Biogenic CH$_4$ Supply Rate Variations

Next, we investigated the relationship between atmospheric $p$CH$_4$ and $\Phi_\uparrow$(CH$_4$) using the photochemical model decoupled from the ecosystem model, to investigate the atmospheric response over a broader range of CH$_4$ flux values (Figure 6). Atmospheric $p$CH$_4$ responded nonlinearly to $\Phi_\uparrow$(CH$_4$) variations for $\Phi_\uparrow$(CH$_4$) values of 3–20 Tmol C yr$^{-1}$ depending on atmospheric $p$CO$_2$. For example, for atmospheric $p$CO$_2$ equal to 0.1 bar, the atmospheric $p$CH$_4$ response becomes nonlinear for $\Phi_\uparrow$(CH$_4$) values over approximately 6 Tmol C yr$^{-1}$ (Figure 6a). To identify the main controlling factors for atmospheric $p$CH$_4$, we separated the chemical reactions consuming atmospheric CH$_4$ (Figure 6d, example for the case of $p$CO$_2$ = 0.1 bar). For $\Phi_\uparrow$(CH$_4$) values lower than a threshold value of approximately 8 Tmol C yr$^{-1}$, reaction with OH radicals is the nearly exclusive atmospheric CH$_4$ consumption process. For higher flux values, photodissociation becomes dominant (Figure 6d).

For $\Phi_\uparrow$(CH$_4$) values lower than the threshold value, the CH$_4$–OH reaction rate is constrained by the CH$_4$ supply rate to the troposphere. Once the residing atmospheric CH$_4$ is consumed by reaction with OH, the relationship for CH$_4$ between its biogenic flux and volume mixing ratio ($f_{CH4}$) is approximated by:

$$f_{CH4} \simeq \frac{\Phi_\uparrow(CH_4)}{\int_0^{z_{crit}} K \cdot n_{OH}(z) \cdot n_{atm}(z) \cdot dz}, \qquad (6)$$

where $n_{atm}(z)$ and $n_{OH}(z)$ are the air and OH number densities at altitude $z$, $K$ is the reaction rate constant, and $z_{crit}$ is the maximum altitude below which CH$_4$ is well mixed ($z_{crit} \sim 60$ km). If $n_{OH}$ is independent from $\Phi_\uparrow$(CH$_4$), then $f_{CH4}$ and $\Phi_\uparrow$(CH$_4$) should be linearly correlated.

When $\Phi_\uparrow$(CH$_4$) approaches the threshold value, the CH$_4$–OH reaction becomes less sensitive to $\Phi_\uparrow$(CH$_4$) increases because the reaction rate is constrained by the availability of OH radicals in the troposphere (Figure 6d). As a result of consumption by biogenic CH$_4$, the OH surface mixing ratio decreases sharply. This confirms that the nonlinearity of the atmospheric $p$CH$_4$ response is directly related to OH and CH$_4$ abundance variations when $\Phi_\uparrow$(CH$_4$) is above the threshold (Figure 6). Excess biogenic CH$_4$ that was not consumed in the troposphere is transported into the mesosphere and thermosphere, where it is photodissociated by strong solar ultraviolet (UV) radiation. This transition between chemical reactions regulating the atmospheric $p$CH$_4$ explains its nonlinear response to high biogenic CH$_4$ supply fluxes into the atmosphere.

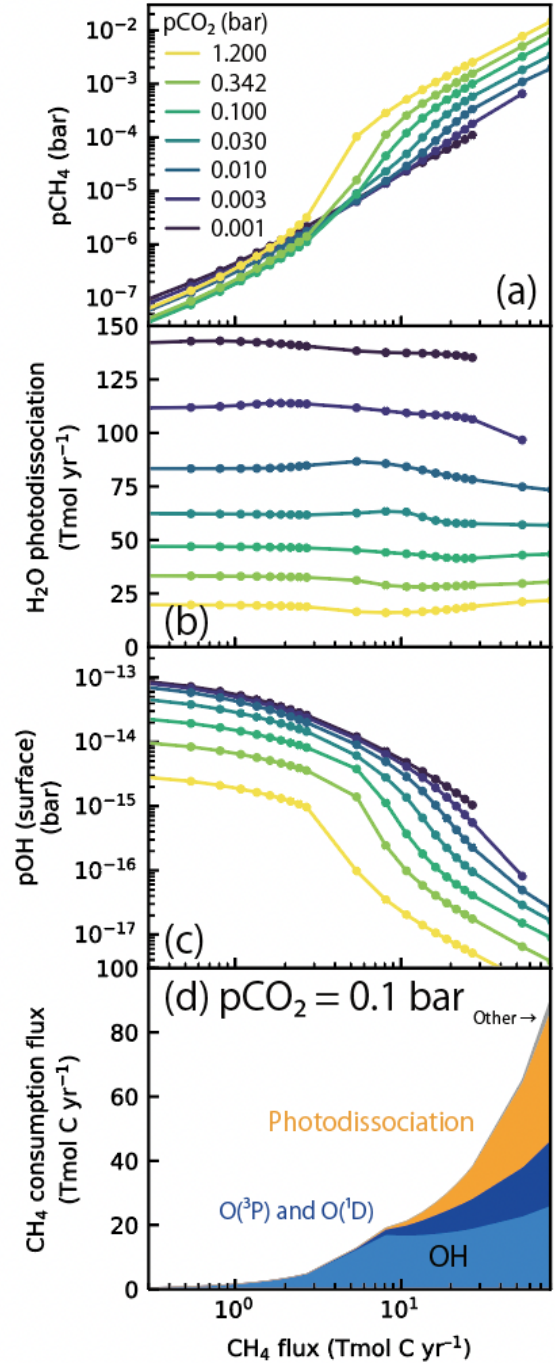

Figure 6. Changes in atmospheric $pCH_4$ (a), photodissociation rate of $H_2O$ (b), partial pressure of OH at the surface (c), and the consumption pathway of the biogenic $CH_4$ flux to the atmosphere (d). Calculations are conducted with $pCO_2$ of 0.001, 0.003, 0.01, 0.03, 0.1, 0.342, and 1.2 bar (*ExpCH4flux* in Table 3). The line colors represent the atmospheric $pCO_2$. The result shown in (d) is for atmospheric $pCO_2$ of 0.1 bar.

3.4. Effect of Atmospheric $pCO_2$ Variations on $CH_4$ Amplification

The next set of simulations was designed to evaluate the dependence of atmospheric $CH_4$ abundances on $pCO_2$ (Figure 6, color-coding of curves in Panels a–c). The nonlinearity of the atmospheric $pCH_4$ response, explained in Section 3.3, increased with increasing atmospheric $pCO_2$, as illustrated in Figure 6, for example, for a $pCO_2$ value of 1.2 bar (Panel (a), yellow curve). Conversely, for a $pCO_2$ value of 0.001 bar, atmospheric $pCH_4$ depends nearly linearly on $\Phi_\uparrow(CH_4)$. We attribute this to the influence of atmospheric $pCO_2$ on tropospheric availability of OH radicals (Figure B1b). The source of tropospheric OH radicals is $H_2O$ photodissociation (Figure 5h). Below the tropopause (approximately 10 km), the $H_2O$ abundance is controlled by the adiabatic temperature gradient, which was fixed in these simulations. The $H_2O$ photodissociation rate ($F_{phot,H2O}$) is expressed as:

$$F_{phot,H2O} = \int_z n_{H2O}(z) \cdot \int_\lambda k_p(z,\lambda) d\lambda dz, \qquad (7)$$

where $n_{H2O}(z)$ is the $H_2O$ number density at altitude $z$ and $k_p(\lambda, z)$ is the photodissociation rate constant for $H_2O$, defined as:

$$k_p(\lambda, z) = \sigma(\lambda) \cdot \Phi(\lambda) \cdot F(\lambda, z), \qquad (8)$$

where $\sigma(\lambda)$ is the $H_2O$ absorption cross-section at wavelength $\lambda$, $\Phi(\lambda)$ is the quantum yield for $H_2O$ photodissociation ($\Phi(\lambda) = 1$), and $F(\lambda, z)$ is the stellar irradiance.

Calculated wavelength-dependent $k_p$ at the tropopause is plotted as a function of $pCO_2$ (color-coded curves) in Figure 7b. For an atmospheric $pCO_2$ value of 0.001 bar, the $k_p$ value at the tropopause peaked at ~170–210 nm, indicating that tropospheric $H_2O$ photodissociation is mainly controlled by solar irradiance in this wavelength range. When steady-state $pCO_2$ increases, the lower wavelength range limit increases, because of the shielding of the tropospheric $H_2O$ by $CO_2$. When $pCO_2$ reaches 0.03 bar, $CO_2$ almost completely absorbs the UV radiations at $CO_2$ absorption bands. For higher values of atmospheric $p$CO$_2$, H$_2$O photodissociation only occurs within ~200–210 nm (Figure 7c). Therefore, the UV flux at 200–210 nm is the dominant controlling factor for H$_2$O photodissociation, indicating that solar irradiance outside of this narrow spectral range does not strongly affect tropospheric H$_2$O photodissociation for $p$CO$_2$ values higher than 0.03 bar (Figure 7), hence that the H$_2$O photodissociation rate is inversely related to atmospheric $p$CO$_2$ (Figure 6b). This minimized OH availability in the troposphere and contributing to atmospheric CH$_4$ amplification, by inducing a nonlinear $p$CH$_4$ increase when atmospheric $p$CO$_2$ is sufficiently high to completely absorb solar radiation at 170–200 nm.

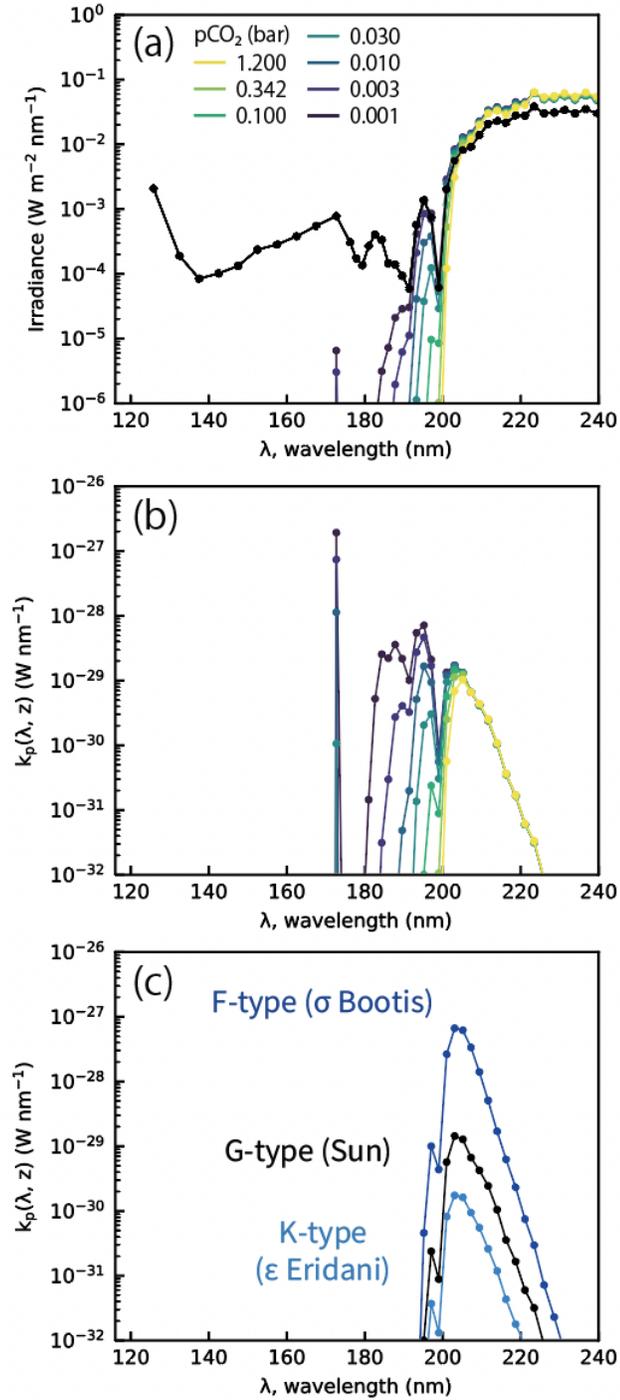

Figure 7. Irradiance at the top of the atmosphere (black) and at the troposphere (a) for the case of the early Earth (*ExpCH4flux* in Table 3). A rate constant for the photodissociation of $H_2O$ at the tropopause (approximately 10 km) (b and c). The line colors represent the results with different atmospheric $pCO_2$ in (a) and (b), and the results using the spectrum of F-, G-, and K-type stars (blue, black, and light blue lines) (*ExpCH4fluxF2V*, *ExpCH4fluxG2V*, and *ExpCH4fluxK2V* in Table 3) in (c).

3.5. Effect of Stellar Spectra on CH$_4$ Amplification

Finally, we conducted several sets of simulations to analyze the dependence of $p$CH$_4$ on $\Phi_\uparrow$(CH$_4$), as in Section 3.3, but under stellar irradiance conditions representative of three central star types (Section 2.2, Figure 2, and Table 3). Results are illustrated in Figure 8. The simulated atmospheric $p$CH$_4$ was lower with the F-type star spectrum than with the G-type star spectrum, but higher with the K-type star spectrum. We attribute this to stellar irradiance differences between F-, G-, and K-type stars. Specifically, the stellar irradiance at 200–210 nm, critical for H$_2$O photodissociation under high $p$CO$_2$ conditions (Section 3.4), was 10 times weaker and 40 times stronger (Figure 2) for the K-type and F-type stars, respectively, than for the G-type star (Sun). Such large irradiance discrepancies would critically affect the H$_2$O photodissociation rate, hence also atmospheric $p$CH$_4$. To confirm this, we repeated the G-type star simulation (with an atmospheric $p$CO$_2$ of 0.1 bar) but scaled the irradiance at approximately 200–210 nm relatively to its nominal value (Figure 2). For weaker stellar irradiances (0.5 and 0.1 times that of the G-type star), atmospheric $p$CH$_4$ increased and the $\Phi_\uparrow$(CH$_4$) threshold value (nominally 8 Tmol C yr$^{-1}$, Section 3.3) decreased. This mechanism is also driven by variations in the tropospheric H$_2$O photodissociation rate, as discussed for early Earth in Section 3.4. Opposite variations are obtained for stronger stellar irradiances at 200–210 nm (2, 10, and 40 times that of the G-type star).

Simulation results for the 0.1 and 40 times scaling were comparable with those for the K- and F-type stars, respectively (Figure 8). Results for the K-type star yielded slightly higher atmospheric $p$CH$_4$ than for the G-type star with 0.1 times scaling, possibly because of the spectral properties of K-type stars at wavelengths shorter than 200 nm that potentially affect the CO$_2$ and CH$_4$ photodissociation rates. For low atmospheric $p$CO$_2$ (0.001 and 0.01 bar), the lower limit of the $k_\mathrm{p}$ peak wavelength range was 170 nm both for F- and K-type stars, similarly to the G-type star simulations (Figure B2b and B2c). This similarity suggests that $H_2O$ photodissociation is the primary control mechanism for atmospheric $pCH_4$, within a wide range of $pCO_2$ values, in the atmospheres of terrestrial exoplanets orbiting Sun-like stars (Figures B2–B3).

In the final set of simulations, we set the top-of-atmosphere solar irradiance by consecutively applying a global scaling factor (0.2–1.2 times the current Sun value) to the solar irradiance spectrum at all wavelengths, to simulate a range of conditions between the outer and inner edges of a habitable zone. Higher atmospheric $pCH_4$ (nearly one order of magnitude higher than on Earth, Figure 9) was calculated for the solar irradiance of 0.3 times the current value, representative of an exoplanet located farther from its central star near the outer edge of the habitable zone under the corresponding $\Phi_\uparrow(CH_4)$ value of ~9.1 Tmol C yr$^{-1}$. By reducing the tropospheric $H_2O$ photodissociation rate, the weaker stellar irradiance was directly responsible for the high simulated $pCH_4$, possibly indicating that planets orbiting near the outer edge of the habitable zone might be more suitably located to achieve high atmospheric $pCH_4$.

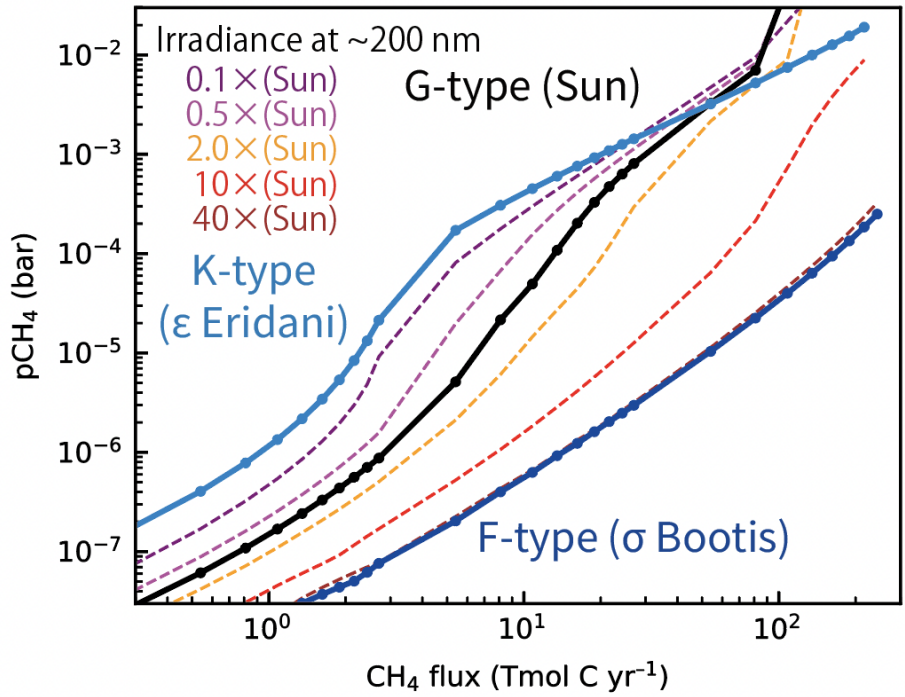

Figure 8. Responses of the atmospheric $p$CH$_4$ to biogenic CH$_4$ flux for the case of F-, G-, and K-type stars (blue, black, and light blue solid lines with dots) (*ExpCH4fluxF2V*, *ExpCH4fluxG2V*, and *ExpCH4fluxK2V* in Table 3). The dotted lines represent the results of the case of the G-type star with the scaled intensity of the spectra at 200–210 nm (0.1, 0.5, 2, 10, and 40 times for purple, light purple, orange, red, and brown lines, respectively). (*ExpCH4flux200nm* in Table 3).

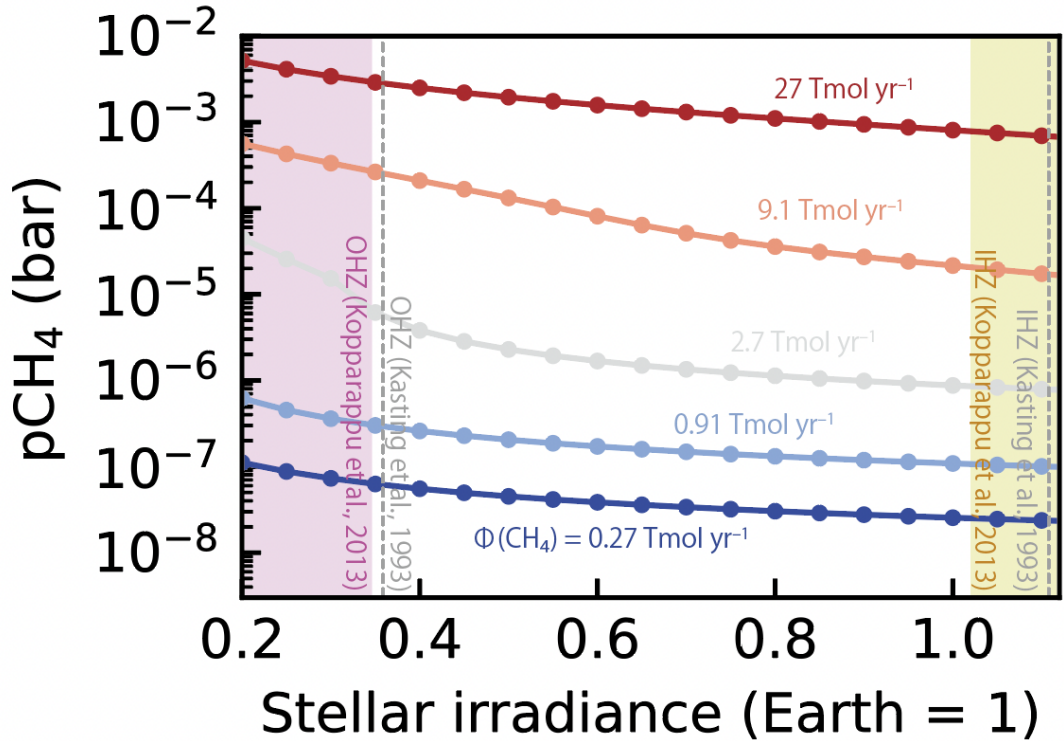

Figure 9. Responses of the atmospheric $p$CH$_4$ to biogenic CH$_4$ flux for the case of G-type star to different stellar irradiance (*ExpIrradiance* in Table 3). Different lines represent the different biogenic CH$_4$ fluxes ($1\times10^9$, $3\times10^9$, $1\times10^{10}$, $3\times10^{10}$, and $1\times10^{11}$ cm$^{-2}$ s$^{-1}$ for blue, light blue, gray, light red, and red lines with dots, respectively). The vertical dashed lines represent the inner and outer edges of the habitable zone (IHZ and OHZ, respectively) estimated by Kasting et al. (1993). The yellow- and purple-hatched regions represent the condition inside the IHZ and the outside the OHZ estimated by Kopparapu et al. (2013), respectively.

4. DISCUSSION

In this study, we evaluated the relationships between flux of reducing power to the ocean–atmosphere system ($H_2$ and CO outgassing fluxes and Fe(II) supply flux), $\Phi_\uparrow(CH_4)$ to the atmosphere, and atmospheric $pCH_4$, by applying our coupled atmospheric photochemistry–primitive marine microbial ecosystem model to the early Earth. Our results demonstrate that CO-C activities are a major factor relating an influx of reducing power to $\Phi_\uparrow(CH_4)$. The effect of a nonlinear increase in atmospheric CO is further enhanced by a decrease in the OH concentration. Moreover, we showed that the tropospheric OH availability is primarily controlled by $H_2O$ photodissociation and, in turn, controls atmospheric $pCH_4$. Consequently, atmospheric $pCH_4$ is affected by atmospheric $pCO_2$ variations via changing $H_2O$ photodissociation rate and the availability of OH radicals. More specifically, the solar irradiance at ~200–210 nm is the primarily controlling factor for the $H_2O$ photodissociation when atmospheric $pCO_2$ exceeds ~0.03 bar. It is expected that atmospheric $pCO_2$ should show a long-term decreasing trend, explained by adjustments of the carbonate-silicate geochemical cycle to a decrease in the $CO_2$ outgassing rate from interior of the Earth and to increases in solar luminosity and land fraction (Walker et al. 1981; Tajika and Matsui 1992; Kasting 1993; Krissansen-Totton et al. 2018; Kadoya and Tajika 2019; Kadoya et al. 2020; Watanabe and Tajika 2021; Watanabe et al. 2023a). Accordingly, nonlinear atmospheric $pCH_4$ amplification likely occurred more easily during the early Archean after the emergence of life. Additionally, early-Earth outgassing rates for reducing gases were likely higher than their current value (Holland 1984; Krissansen-Totton et al. 2018, 2021). Such conditions are suitable for nonlinear amplification of atmospheric $CH_4$ on early Earth that probably occurred immediately after a primitive marine biosphere emerged. However, although OH availability is also affected by tropospheric temperature, which demonstrably influences tropospheric $H_2O$ concentrations, this temperature effect was not considered in this study; instead, a fixed atmospheric temperature vertical profile was used. When atmospheric $pCO_2$ is high, surface temperature increases and more OH radicals than simulated in this study are produced in the troposphere. Therefore, the results with our high-$pCO_2$ analysis represent an upper limit to the estimated atmospheric $pCH_4$ values. Nevertheless, we do not expect a strong effect on atmospheric $pCH_4$ because early-Earth climatic conditions would have been regulated by the carbonate-silicate geochemical cycle.

We also explored the effect of stellar irradiance on the atmospheric $pCH_4$ variations for Earth-like inhabited exoplanets, both depending on their location inside the habitable zone and on the central star. Our results show that exoplanetary atmospheric $pCH_4$ variations are also strongly controlled by the stellar spectrum intensity. Inhabited exoplanets around K-type stars would exhibit markedly higher atmospheric $pCH_4$ than F- or G-type stars, providing optimal conditions to search for life (Grenfell et al. 2007; Arney 2019). Conversely, the strong stellar irradiance provided by F-type stars would likely prevent a strong buildup of atmospheric $pCH_4$ on their planets.

Furthermore, we demonstrated that inhabited exoplanets orbiting near the outer edge of the habitable zone are more likely to exhibit high atmospheric $pCH_4$ because the correspondingly weaker stellar irradiance would enhance atmospheric $pCO_2$ through the carbonate-silicate geochemical cycle (Kadoya and Tajika 2014, 2019), inducing an increase in atmospheric $pCH_4$. In such a case, tropospheric $H_2O$ photodissociation and atmospheric $pCH_4$ would be strongly controlled by the 200–210 nm stellar irradiance under high atmospheric $pCO_2$ conditions. This would also provide an ideal setting to produce high atmospheric $pCH_4$ (Grenfell et al. 2007). Planets orbiting Sun-like stars near the outer edge of their habitable zone are less likely to present a false positive O$_2$ biosignature (Meadows 2017). Thus, they are optimally suited to search for extraterrestrial life. Incidentally, high atmospheric $p$CO$_2$ conditions, promoted by the limited stellar irradiance at the outer edge of the habitable zone, may also affect the exoplanetary atmospheric composition through the formation of CO$_2$ clouds in the polar regions (Kasting 1991; Pierrehumbert and Erlick 1998).

Conversely, near the inner edge of the habitable zone (solar irradiance of 1.01–1.10 times that of the current Sun) (Kasting et al. 1993; Kopparapu et al. 2013), atmospheric $p$CH$_4$ is comparable with standard G-type star irradiance, but marginally lower. In this case, steady-state atmospheric $p$CH$_4$ should be considerably smaller (Grenfell et al. 2007), because the higher surface temperature would yield a larger tropospheric H$_2$O abundance than that inferred from the input profile adopted in this study. Therefore, a comprehensive approach combining a global carbon cycle model, a climate model, a photochemical model, and a marine microbial ecosystem model is essential to better constraining atmospheric $p$CH$_4$ in exoplanetary atmospheres orbiting within the habitable zone of Sun-like stars. For exoplanets orbiting the M-type stars, which is not explored in this study, spectra of the stellar irradiance at ultraviolet wavelengths would be different from those of Sun-like stars. The controlling factor of atmospheric $p$CH$_4$ on exoplanets around M-type stars should also be investigated in future studies.

## 5. CONCLUSION

In this study, we investigated controlling factors influencing the atmospheric CH$_4$ abundance of early Earth and of potentially inhabited Earth-like exoplanets orbiting Sun-like stars (F-, G-, and K-type), using a 1D atmospheric photochemical model and a primitive marine microbial ecosystem model. Nonlinear amplification of the early-Earth $\Phi_{\uparrow}(CH_4)$ into the atmosphere is explained primarily by an amplification of the atmospheric CO cycle and by the activity of CO-C. The nonlinear atmospheric $pCH_4$ response to $\Phi_{\uparrow}(CH_4)$ variations is caused by a decrease in the $H_2O$ photodissociation rate. Atmospheric $CH_4$ abundances on inhabited Earth-like exoplanets orbiting Sun-like stars are mainly controlled by $\Phi_{\uparrow}(CH_4)$ and stellar irradiance. Potentially inhabited Earth-like exoplanets orbiting K-type stars near the outer edge of their habitable zone would be a suitable candidate for detection of $CH_4$ biosignatures.

## 6. Acknowledgments

This work is supported by Grant-in-aid for JSPS Research Fellow Number 20J12951 (YW). We thank Kazumi Ozaki for fruitful discussions. We thank Eric Dupuy, Ph.D., from Edanz (https://jp.edanz.com/ac) for editing a draft of this manuscript.

## 7. Competing interest

The authors declare no competing interest.

APPENDIX A

The absorption cross-section of $CO_2$ and $H_2O$ employed in this study is shown in Figure A1.

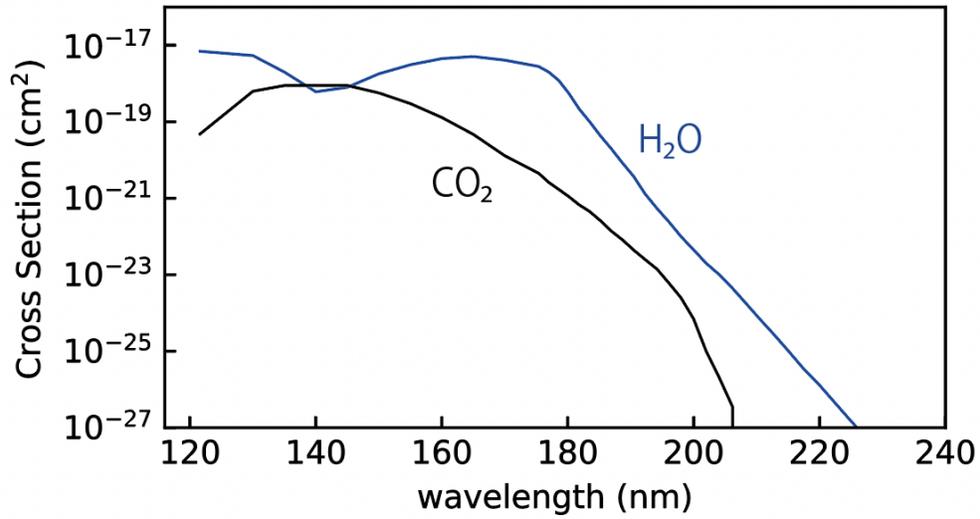

Figure A1. Absorption cross-section of $CO_2$ and $H_2O$ adopted in this study.

APPENDIX B

The vertical profiles of atmospheric species and reaction rates with respect to different atmospheric $pCO_2$ are shown in Figure B1. The rate constants for the photodissociation of $H_2O$ at the tropopause (10 km) and atmospheric $pCH_4$ for different atmospheric $pCO_2$ values and central stars are shown in Figures B2 and B3, respectively.

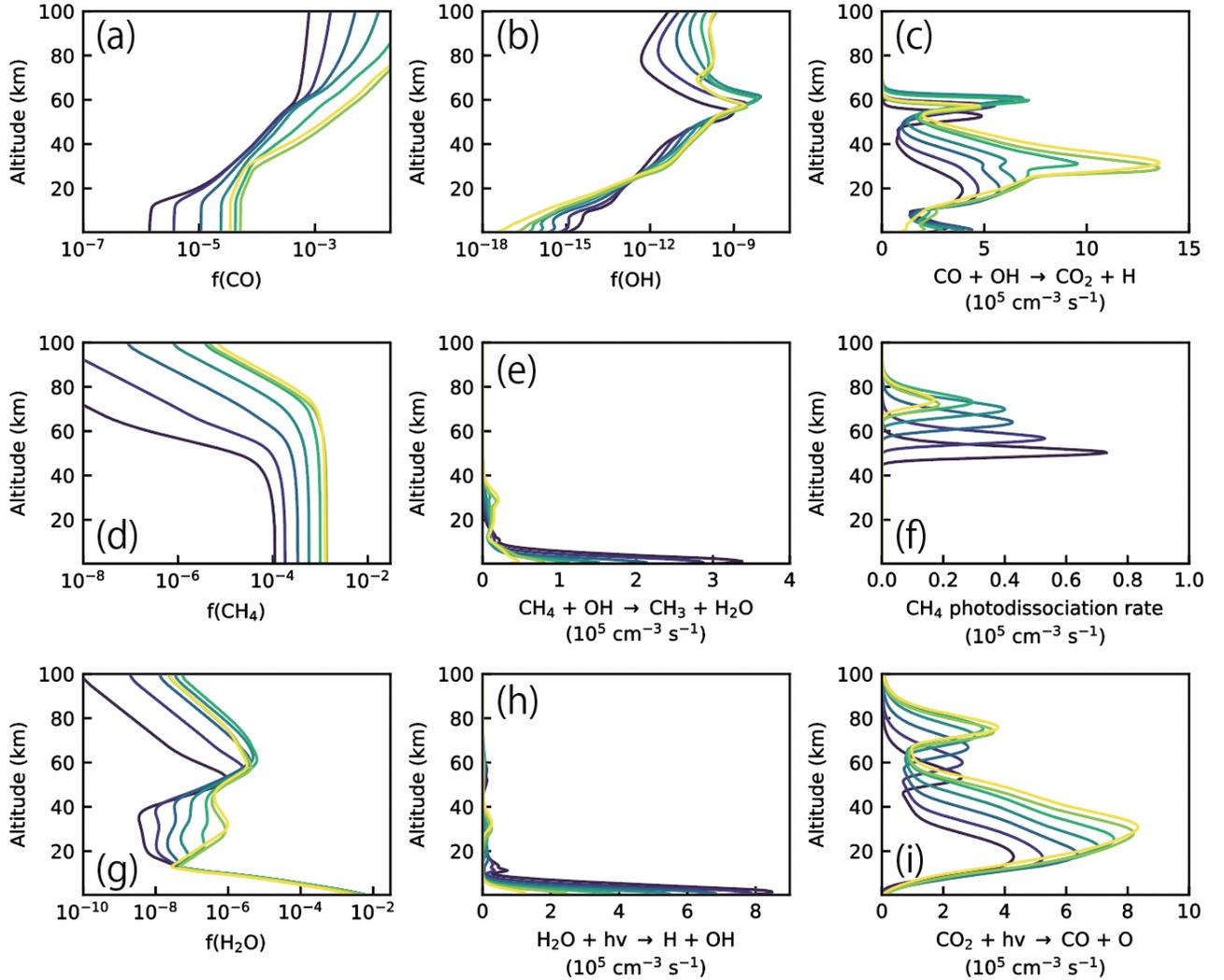

Figure B1. Vertical profiles of the atmospheric components and the reaction rates. Vertical profiles of (a) CO mixing ratio, (b) OH radical mixing ratio, (c) reaction rate of $CO + OH \rightarrow CO_2 + H$, (d) $CH_4$ mixing ratio, reaction rate of (e) $CH_4 + OH \rightarrow CH_3 + H_2O$, (f) photodissociation rate of $CH_4$, (g) $H_2O$ mixing ratio, (h) photodissociation rate of $H_2O$, and (i) photodissociation rate of $CO_2$. Calculations are conducted with different atmospheric $pCO_2$ and a fixed $CH_4$ flux (2.7 Tmol C yr$^{-1}$) (*ExpCH4flux* in Table 3). The line colors represent atmospheric $pCO_2$ (0.001, 0.003, 0.01, 0.03, 0.1, 0.342, and 1.2 bar).

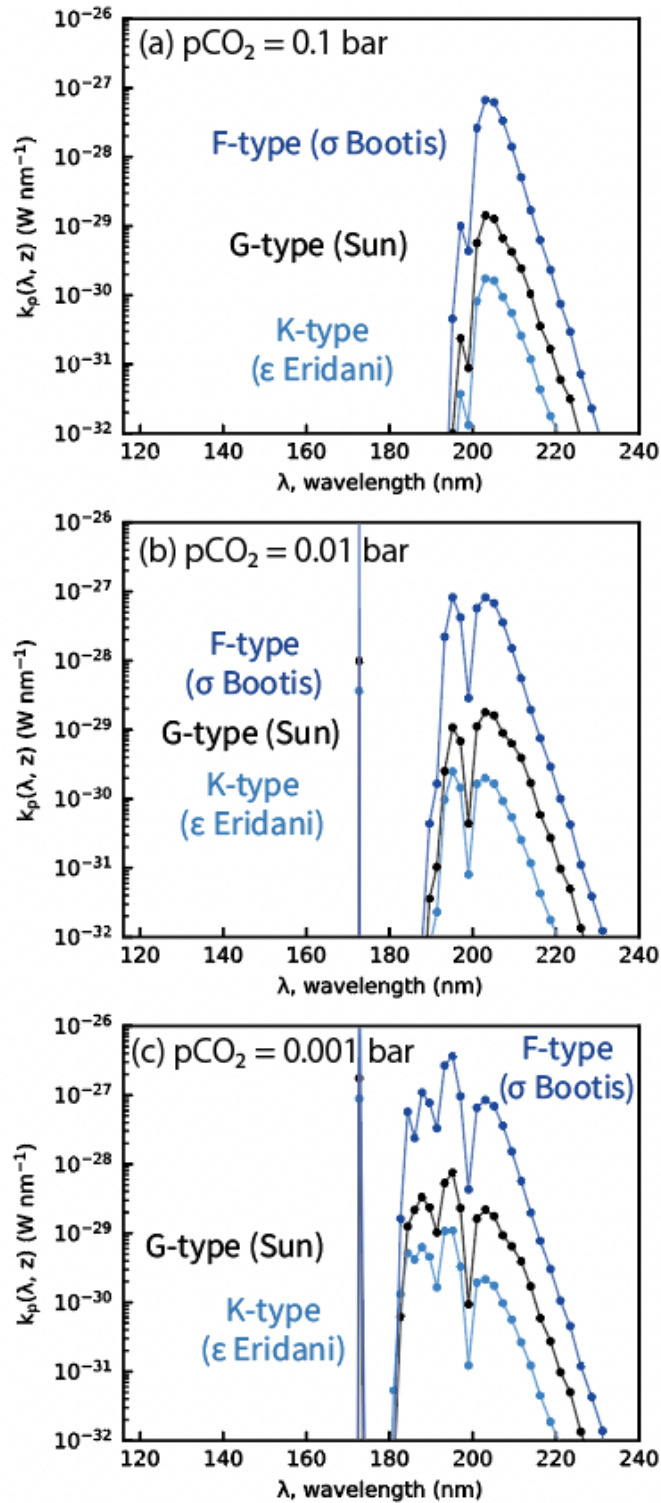

Figure B2. Rate constants for the photodissociation of $H_2O$ at the tropopause (approximately 10 km) for different atmospheric $pCO_2$ (0.1, 0.01, and 0.001 bar for a, b, and c, respectively). The line colors represent the results using the spectrum of F-, G-, and K-type stars (blue, black, and light blue lines) (*ExpCH4fluxF2V*, *ExpCH4fluxG2V*, and *ExpCH4fluxK2V* in Table 3).

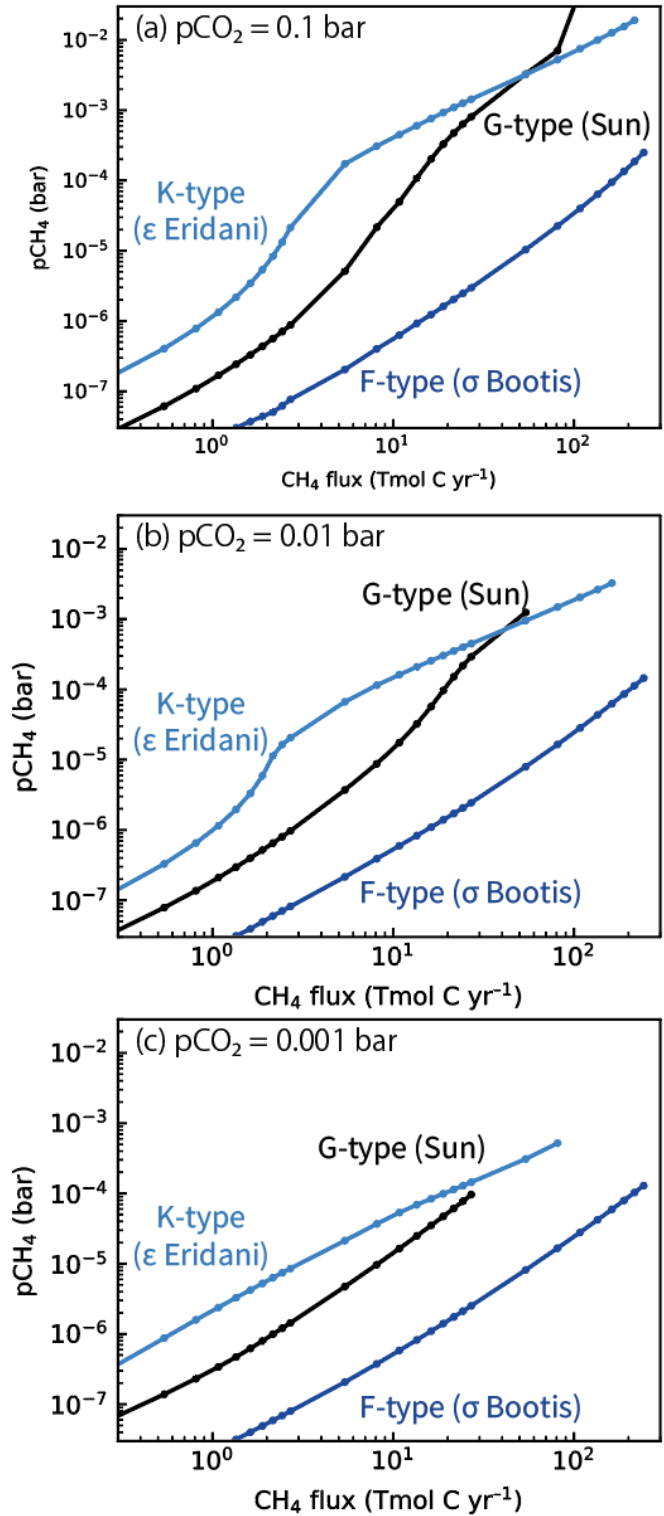

Figure B3. Responses of the atmospheric $p$CH$_4$ to biogenic CH$_4$ flux for the case of F-, G-, and K-type stars (blue, black, and light blue solid lines with dots) (*ExpCH4fluxF2V*, *ExpCH4fluxG2V*, and *ExpCH4fluxK2V* in Table 3) calculated with the atmospheric $p$CO$_2$ of 0.1, 0.01, and 0.001 bar (a, b, and c, respectively).